\documentclass[a4paper]{article}

\usepackage{INTERSPEECH2022}
\usepackage{amsmath,graphicx}
\usepackage{comment}
\usepackage{cite}
\usepackage{caption, color}
\usepackage{adjustbox}

\usepackage[labelformat=simple]{subcaption}

\title{Impact of Acoustic Event Tagging on Scene Classification in a Multi-Task Learning Framework}

\name{Rahil Parikh$^1$, Harshavardhan Sundar$^2$, Ming Sun$^2$, Chao Wang$^2$, Spyros Matsoukas$^2$ \vspace*{-3mm}}

\address{$^1$ University of Maryland,
	College Park, MD, USA \\
	$^2$Alexa AI,
	Amazon.com Inc,
	Cambridge, MA, USA}

\email{rahil@umd.edu, sundarhs@amazon.com}

\begin{document}

\maketitle
\begin{abstract}
Acoustic events are sounds with well-defined spectro-temporal characteristics which can be associated with the physical objects generating them. Acoustic scenes are collections of such acoustic events in no specific temporal order. Given this natural linkage between events and scenes, a common belief is that the ability to classify events must help in the classification of scenes. This has led to several efforts attempting to do well on Acoustic Event Tagging (AET) and Acoustic Scene Classification (ASC) using a multi-task network. However, in these efforts, improvement in one task does not guarantee an improvement in the other, suggesting a tension between ASC and AET. It is unclear if improvements in AET translates to improvements in ASC. We explore this conundrum through an extensive empirical study and show that under certain conditions, using AET as an auxiliary task in the multi-task network consistently improves ASC performance. Additionally, ASC performance further improves with the AET data-set size and is not sensitive to the choice of events or the number of events in the AET data-set. We conclude that this improvement in ASC performance comes from the regularization effect of using AET and not from the network’s improved ability to discern between acoustic events.

\end{abstract}

\section{Introduction}
\label{sec:intro}
Computational Environmental Audio Analysis \cite{virtanen2018computational} aims to extract and interpret information related to the environment from which an acoustic signal is recorded. Two major sub-classes of problems in this field are Acoustic Event Tagging (AET) \cite{wang2019comparison, kao2020joint, fonseca2018general, kong2017joint} and Acoustic Scene Classification (ASC) \cite{barchiesi2015acoustic, bae2016acoustic, battaglino2016acoustic, geiger2013large}. AET is a multi-label classification problem which involves classifying an audio sample into one or more predefined events. ASC is a multi-class classification problem in which an audio sample is classified into one of the predefined acoustic scenes that represents the environment from which the audio sample was recorded. Acoustic events represent information at lower levels of abstraction with clear time-frequency patterns such as `car engine', `dog-bark', etc., while scenes are collection of acoustic events in no specific temporal order and represent information at higher levels of abstraction such as `street traffic' and `urban park'. 

Studies in auditory perception and cognition suggest that humans leverage event information for scene classification \cite{guastavino2020current}. For instance, knowledge of the event `jet-engine' helps classify a given acoustic scene as `airport' instead of `shopping mall'. This has inspired several researchers to integrate ASC with AET or Acoustic Event Detection (AED) \cite{jung2020acoustic, tonami2019joint, bear2019towards, jung2021dcasenet, zhang2019cross}. These works are grounded on the assumption that a model's capability to classify or detect events (i.e. to perform AET or AED) must help in its ability to classify scenes (i.e. to perfrom ASC). Various frameworks and model architectures have been implemented in an attempt to use the AET/AED task to improve ASC performance. In \cite{zhang2019cross} the authors explore pre-training a network on the AET task and then fine-tuning it to perform ASC.
Jointly training for AED/AET and ASC using a multi-task learning framework has been explored in \cite{jung2020acoustic, tonami2019joint, bear2019towards, jung2021dcasenet}. These works employ a common encoder and task-specific decoders for ASC and AET. Here, the network is trained on a linear combination of the individual task losses, thereby presenting multiple hyper-parameters that need to be tuned. In the absence of an existing data set with both event and scene annotations, most works use task-specific individual data sets from the DCASE Challenges to train each task. However, in each of these works, researchers have observed that jointly training with AET does not guarantee an improvement in ASC performance. In \cite{jung2021dcasenet}, various multi-task learning model configurations are explored to leverage event and scene cross-information sharing for ASC, AET and AED tasks, despite which, the improvements in performance are often marginal.

The marginal improvements obtained from jointly-training on AET and ASC tasks suggest a `tension' between the two tasks which prevents them from simultaneous improvements. A likely explanation for this tension arises from the limitations of the human annotation process and data collection \cite{abesser2020review, wu2019enhancing}. To elaborate, data sets are commonly annotated with `foreground' sounds, such as \textit{speech}, \textit{distinct musical instruments} etc. These sounds are informative of the acoustic events. In contrast, `background' sounds such as \textit{wind}, \textit{distant home appliances} etc. which are more informative of the scene are seldom annotated. In a model jointly trained to optimize ASC and AET performance, an improvement in the AET performance may come at the cost of the ASC performance since low SNR regions are ignored. Similarly, the ASC data set contains several un-annotated events for each scene which may or may not be present in the AET data set. Jointly-training a network on such data may degrade the AET performance. This tension questions the improvement of the scene classification task when jointly-trained with AET, and suggests the presence of an optimal weighing of the objective function for the AET and ASC tasks to obtain this improvement, if any.

The works in \cite{jung2020acoustic, tonami2019joint, bear2019towards, jung2021dcasenet} are aimed towards improving ASC and AET performance by jointly-training the network on fixed, pre-defined AET and ASC tasks. Instead, in this paper, we probe the tension between AET and ASC tasks in a multi-task learning framework by fixing the ASC task and analyzing the impact on this fixed ASC task by changing various attributes of the AET task. To facilitate this, we create multiple AET data sets by varying the distribution of event labels, the size of the data set and the number of the event labels and examine its effect on the fixed ASC task.

Through such empirical studies we show that:
\begin{itemize}
    \item Performance improvements in AET task are not in general correlated with improvements in ASC task. With AET and ASC tasks fixed, the performance improvement in AET \textbf{does not} guarantee performance improvement in ASC.
    \item Consistent improvements in ASC (up to 5\% accuracy compared to a competitive baseline) are observed when the multi-task loss function is balanced w.r.t. the contribution of the individual AET and ASC losses. 
    \item Improvement in ASC depends on the AET training data set size and is less sensitive to the number of events being classified or the choice of events used in the AET task.
\end{itemize}

Thus, our empirical studies reveal that improvements in ASC can be attributed to the regularization effect of using AET as an auxiliary task. Interestingly, the regularization effect of using AET to improve ASC is more effective than some of the well known regularization strategies of Mix-up \cite{zhang2017mixup}, Dropout \cite{srivastava14adropout}, and SpecAugment \cite{park2019specaugment}. 

The factors like AET training data size, choice and the number of events used can be used as parameters to control the regularization effect. We emphasize that the motivation of our work is not to improve scene classification performance using a multi-task learning framework, but instead to investigate the impact of AET as an auxiliary task on scene classification with empirical evidence to analyze what controls on the AET task improve ASC performance.

\section{Multi-task Learning for ASC and AET}
\label{sec:methodology}
\begin{figure}[t]
  \centering
    \includegraphics[width=1.0\columnwidth]{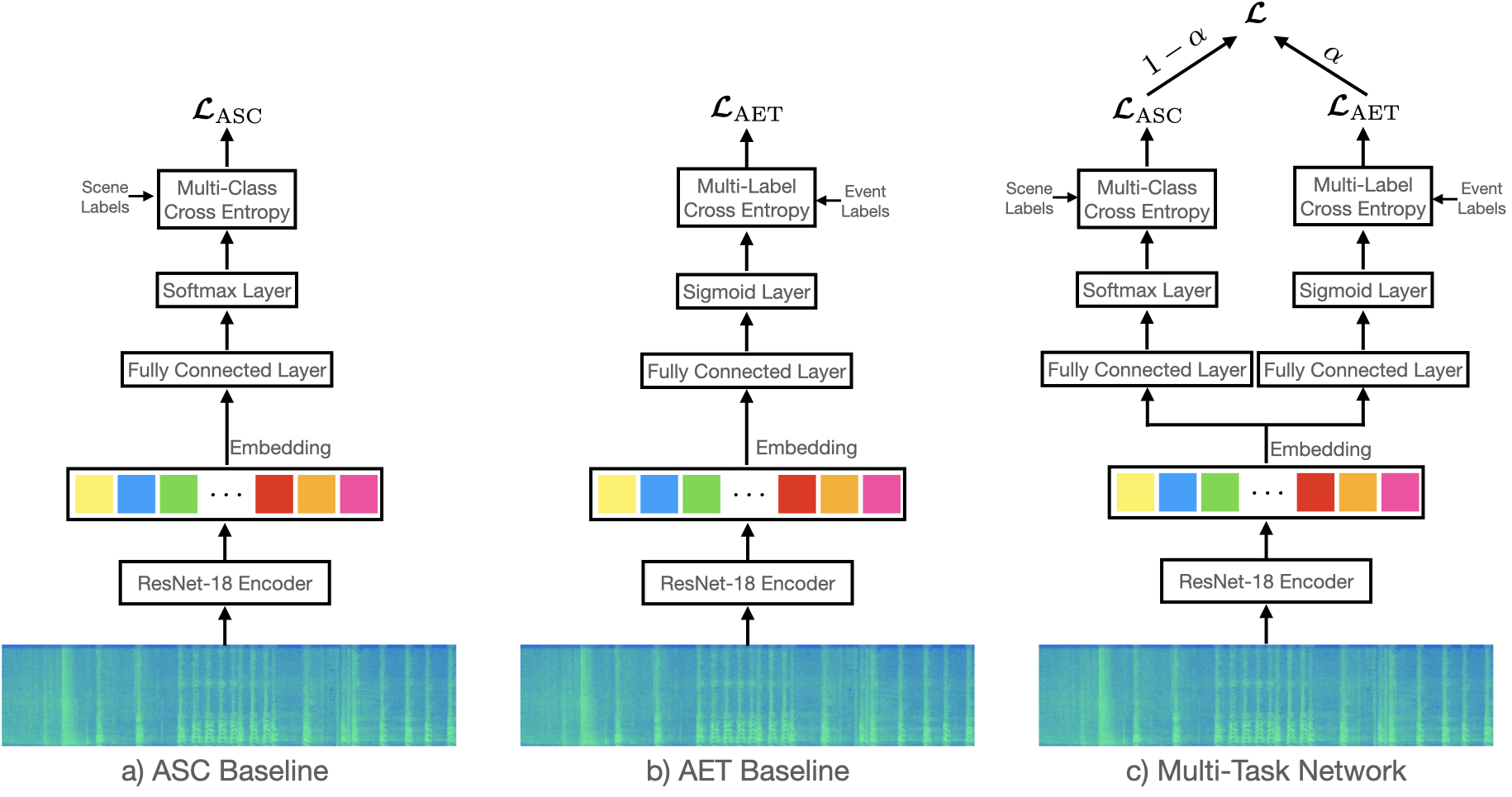}
  \caption{The ASC and AET baselines along with the multi-task network used in this work.}
  \label{fig: model_architecture}
\end{figure}

We jointly train our network to perform ASC and AET using a multi-task learning approach. The two tasks share an encoder that generates an embedding from the input features. This is followed by the ASC head-- a fully-connected layer with softmax activation and an AET head-- a fully-connected layer with sigmoid activation. A ResNet-18 \cite{he2016deep} architecture is used as our common encoder. We use log-mel-spectrograms as input features of $10$s long audios. Figure \ref{fig: model_architecture} shows the individual task specific baselines along with the multi-task network used for analysis. 

The network is trained on the overall loss ($\boldsymbol{\mathcal{L}}$) which is a convex combination of the individual AET and ASC losses parameterized by $\alpha$ as: 
\begin{equation}
    \boldsymbol{\mathcal{L}} = \alpha \cdot \boldsymbol{\mathcal{L}}_{\text{AET}} + (1-\alpha)\cdot \boldsymbol{\mathcal{L}}_{\text{ASC}};~\alpha \in \left[0,1\right]
\label{eqn: total_loss}
\end{equation}
where $\boldsymbol{\mathcal{L}}_{\text{AET}}$ and $\boldsymbol{\mathcal{L}}_{\text{ASC}}$ are the multi-label and multi-class classification losses of AET and ASC respectively. Higher the value assigned to $\alpha$, higher will be the contribution of the AET loss to the overall loss. 
During training, each mini-batch contains a fixed number of samples from the AET and ASC data sets. We experimented with different number of samples in the set $\left\{ 8, 16, 32, 64, 128, 256 \right\}$ per mini-batch for the ASC data set keeping the number of samples from the AET data set fixed at 128. Best results were obtained for a mini-batch size of 192 samples with 64 samples from ASC and 128 samples from AET data sets. We used this setting for all other experiments reported in this paper. We use SpecAugment \cite{park2019specaugment} and MixUp \cite{zhang2017mixup} to augment our training data for regularization.
\section{Experimental Setup} \label{sec: Experiments}
\subsection{Datasets} We use the TUT Urban Acoustic Scenes data set from the DCASE 2018 Task-1 challenge \cite{Mesaros2018_DCASE} for the ASC task. This balanced data set consists of  24 hours of audio from 10 urban scenes. Examples of these scenes are: `airport', `street traffic', `metro', etc. Each sample in the data set has a unique scene corresponding to it. We keep this data set fixed throughout our analysis for comparability of ASC performance.

We sample AudioSet \cite{gemmeke2017audio} to create multiple AET data sets by varying the choice of events in the data set, the size of the data set ($T$) and number of events in the data set ($k$). In its entirety, AudioSet is annotated for 530 event-classes and is 5600 hours long. Each sample in AudioSet may contain multiple event-classes.

Data from the AET data set is used to train the common encoder and the AET head while data from the ASC data set is used to train the encoder and the ASC head.

\subsection{Evaluation Metrics}
We evaluate our models for scene-classification and event-tagging using the metrics followed by the DCASE ASC (Task-1, Editions: 2018-'21) and DCASE AET (Task 2, Editions: 2018-'19) Challenge: \textit{\textbf{Macro-Average Accuracy (Accuracy)}} and \textit{\textbf{Label-Weighted Label-Ranking Average Precision (lwlrap)}} respectively. A higher value suggests better model performance for both metrics.

\subsection{Common Encoder Architecture and Baseline System}
The task specific baselines for ASC and AET are shown in Figure \ref{fig: model_architecture} a) and \ref{fig: model_architecture} b) respectively. The ASC baseline was trained using only the TUT Urban Acoustic Scenes data set with SpecAugment \cite{park2019specaugment} and MixUp \cite{zhang2017mixup} techniques for regularization and $\boldsymbol{\mathcal{L}_{\text{ASC}}}$ as the loss function. We obtain a scene classification accuracy of $68 \%$ with this approach which is comparable to the state-of-art single-channel, single model accuracy for the same data set \cite{mariotti2018exploring}. It is worthwhile to note that, we obtain similar performance with the multi-task network shown in Figure \ref{fig: model_architecture} c) with $\alpha = 0$. While we tried other common architectures such as VGG \cite{simonyan2014very} and CRNN variants \cite{shi2016end}, we observe that the ResNet-18 outperforms them. Therefore, we use ResNet-18 as the common encoder for all our subsequent experiments as discussed in Section \ref{sec:methodology}. 

The AET baseline also employs a ReNet-18 encoder and is trained on several different subsets of AudioSet. The performance of this baseline will be specified based on the specific subset of AudioSet in the following sections. 

\section{Analysing the Impact of AET on ASC}
In this section we analyze the impact of changing the AET task in various ways on the ASC task. We run 4 trials for each experiment, initialized with 4 different random-seeds.
\label{sec:Results}
\subsection{Impact of $\alpha$ on ASC Performance}
\label{Improvement in ASC Using Joint Training with AET}

\begin{figure}
  \centering
  \includegraphics[width=\columnwidth]{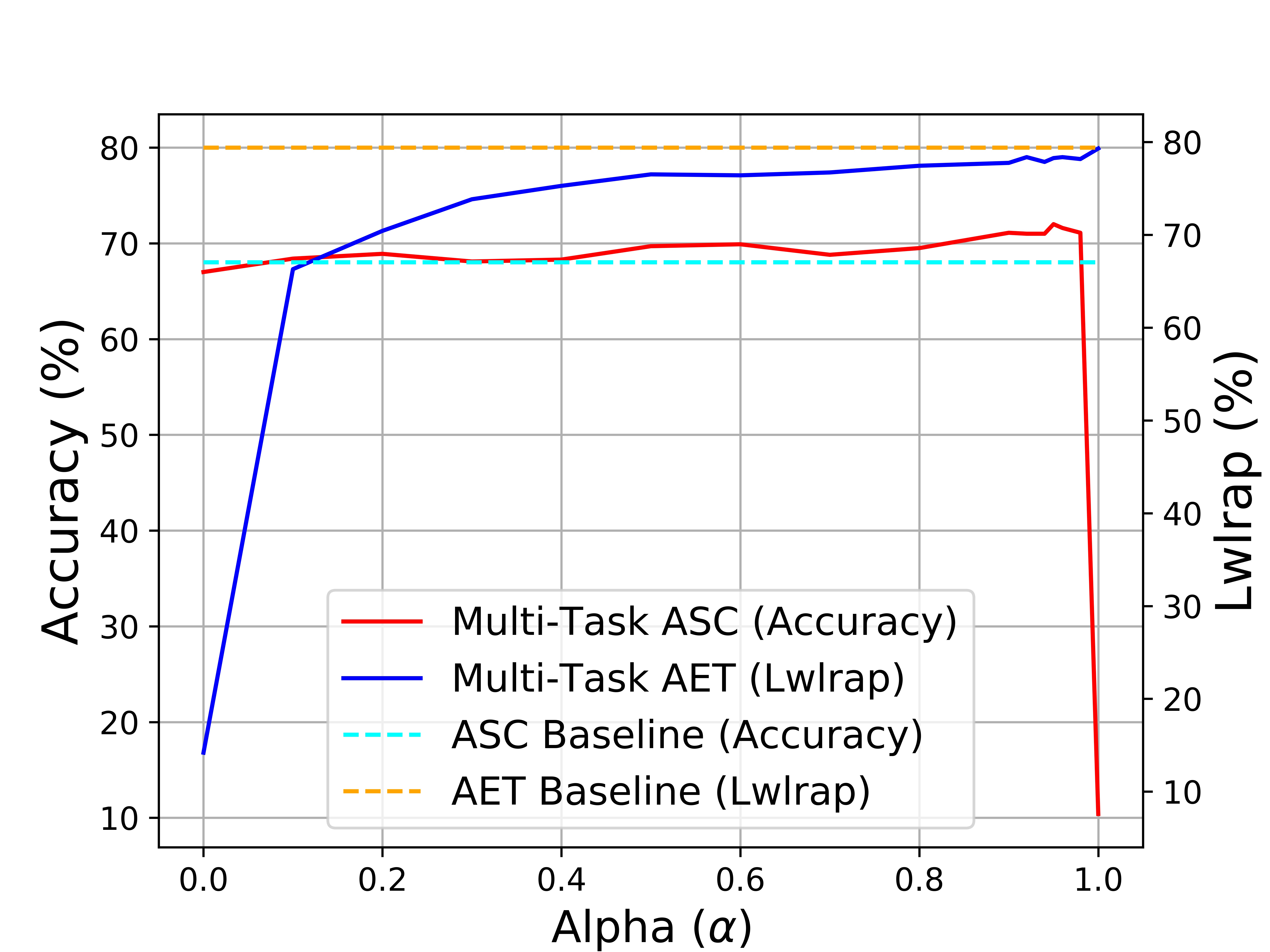}
  \caption{Joint-Training with Different $\alpha$ using the Correlated Event-Set with $k=35$, $T=170$ hours.}
  \label{fig: alpha_sweep}
  \vspace*{-3mm}
\end{figure}

We investigate the impact of changing the relative contribution of the two individual task-specific losses $\boldsymbol{\mathcal{L}}_{\text{AET}}$ and $\boldsymbol{\mathcal{L}}_{\text{ASC}}$ to the overall loss given in (\ref{eqn: total_loss}) by varying $\alpha$. We perform this analysis on the multi-task network shown in Figure \ref{fig: model_architecture} c), using different values of $\alpha$ incremented by 0.1 in the range $\left[0, 0.9\right)$ and by 0.01 in $\left[0.90, 1\right]$. The AET data set is chosen to comprise of 35 randomly selected event classes with a total size of 170 hours. Figure \ref{fig: alpha_sweep} shows the scene classification accuracy and event tagging performance (lwlrap) of the multi-task network along with the performance of the individual task specific baselines. At $\alpha=0$, the model is only trained on $\boldsymbol{\mathcal{L}}_{\text{ASC}}$. The scene accuracy of the multi-task network is thus equivalent to the ASC baseline model while the multi-task network's event tagging performance is equivalent to \textit{random chance}. Similarly, at $\alpha = 1.0$, the multi-task network is only trained on $\boldsymbol{\mathcal{L}_{\text{AET}}}$ and its the event tagging performance is similar to that of the AET baseline while its performance on scene classification is at chance. As $\alpha$ increases from 0 to 1, the event tagging performance (lwlrap) of the multi-task network increases. Strangely, the multi-task network outperforms the ASC baseline in terms of scene classification accuracy for higher values of $\alpha \in \left[0.8, 1.0\right)$. We repeat this experiment on $6$ separate AET data sets sampled from AudioSet each with $35$ randomly chosen events with the total number of hours in the range of $11$ to $360$ hours. In each of these cases, we observe that the performance of the multi-task network improves over the ASC baseline for higher values of $\alpha$. On closer examination, we observe that upon convergence $\boldsymbol{\mathcal{L}}_{\text{AET}}^* \approx \boldsymbol{\mathcal{L}}_{\text{ASC}}^*/20$, where $\boldsymbol{\mathcal{L}}_{\text{AET}}^*$ and $\boldsymbol{\mathcal{L}}_{\text{ASC}}^*$ represent the individual task losses upon convergence. Additionally, the range of $\alpha$ over which we observe consistent improvements of the multi-task network over the ASC baseline, the individual loss terms are nearly equal, i.e., 
\begin{align}
    \alpha . \boldsymbol{\mathcal{L}}_{\text{AET}}^* &\approx \left(1 - \alpha\right).\boldsymbol{\mathcal{L}}_{\text{ASC}}^* \\   
    \therefore \alpha &\approx \frac{\boldsymbol{\mathcal{L}}_{\text{ASC}}^*}{\boldsymbol{\mathcal{L}}_{\text{ASC}}^* + \boldsymbol{\mathcal{L}}_{\text{AET}}^*} \label{eqn:optimal_alpha}
\end{align}

As a simplifier, we use $\boldsymbol{\mathcal{L}}_{\text{ASC}}^*$ and $\boldsymbol{\mathcal{L}}_{\text{AET}}^*$ as converged losses from the training of the respective task specific baselines and fix  $\alpha = 0.95$ for the remainder of the experiments. 

Interestingly, from Figure \ref{fig: alpha_sweep}, we observe that for a multi-task network trained with $\alpha=0.5$ the event lwlrap is around 76\% whereas for a multi-task network trained with $\alpha=0.$, the event lwlrap is around 10\%. Thus, compared to the multi-task network trained with $\alpha=0$, the multi-task network trained with $\alpha=0.5$ is a much better event tagger. However, this improved event tagging ability does not translate to improved scene classification accuracy of the multi-task network. Thus, contrary to the assumptions made in \cite{jung2021dcasenet, tonami2019joint}, our experiments show strong evidence that such multi-task network's improved scene classification does not come for the ability of the network to do better event tagging but rather comes about from the regularization effect of using AET as a challenging auxiliary task. The value of $\alpha$ chosen to balance the two losses contributes to making AET a challenging auxiliary task while the same network is required to also do well on the scene classification task. Additionally, we believe AET is an even more effective regularizer for the ASC tasks than the generally useful SpecAugment \cite{park2019specaugment} and MixUp \cite{zhang2017mixup} techniques. 

\subsection{Impact of Choice of Events in the AET Task on ASC Performance}
\label{sec:Choice of events in the AET task}
To analyze the impact of the choice of event classes in the AET task on ASC, we sample $k$ event-classes with a data set size of 170 hours from AudioSet. The following event-class sampling strategies are explored to evaluate the impact of choosing events used to train the AET task:
\begin{enumerate}
    \item Random Selection: $k$ events are randomly sampled from the set of 530 event-classes present in AudioSet. 
    \item Correlated Event Selection: Each clip from the TUT Urban Acoustic Scenes data set used for the ASC task may contain multiple events. However, these events are not annotated as these clips are only associated with scene labels. We wish to train the AET task on the $k$ events that are most likely to be present in this data set. To predict these events in the ASC data set, we use a trained fine-grained event tagger. This tagger is a DenseNet \cite{huang2018densely} trained on over 3.58M samples from the Amazon Instant Video (AIV) data set and Amazon Alexa's internal production data set for 203 classes. On AIV and the internal production data set, this classifier has a Top-1 accuracy of 34\% and 99.98\% respectively, and a Top-5 accuracy of 65\% and 99.99\% respectively. We choose the $k$ event-classes in AudioSet with the highest prediction scores since they are most-likely to be present in ASC data. Examples of such events are `speech', `traffic noises', `footsteps', `jet-engine' etc. 
    \item Uncorrelated Event Selection: To sample the $k$ events least likely to be present in the TUT Urban Acoustic Scenes data set, we use the above-mentioned event tagger and choose the $k$ events in AudioSet with the lowest prediction scores on the ASC data set. These events are least-likely to appear in ASC data and include events such as `roaring cats', `orchestra', `gunshots', etc. 
\end{enumerate}
Our AET data sets thus consists of a subset of AudioSet data corresponding to $k$ classes, sampled using either of the above-mentioned strategies.

\begin{figure}
  \centering
  \begin{subfigure}[l]{0.23\textwidth}
  \includegraphics[width=1.7in]{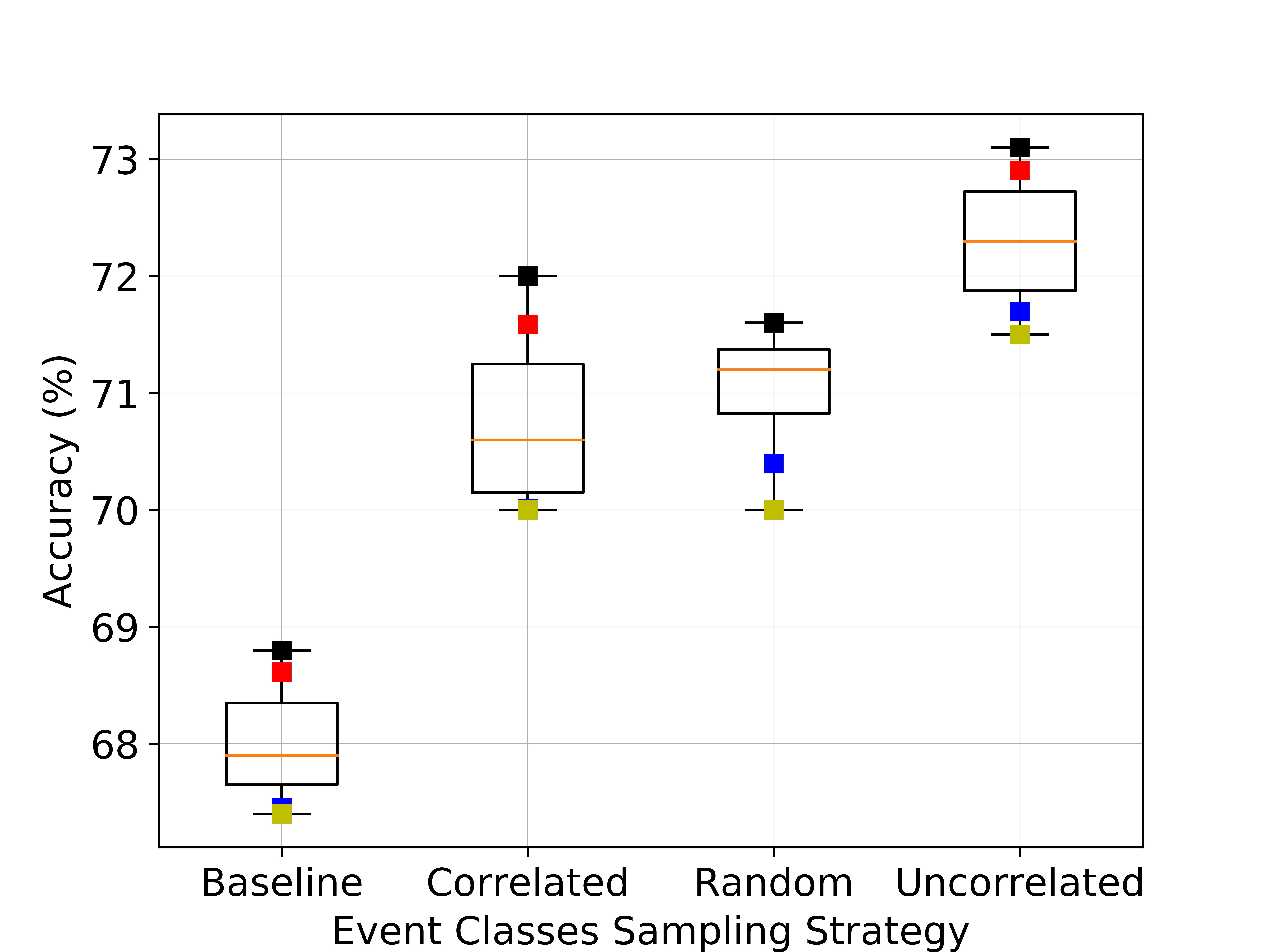}
  \caption{}
  \label{fig: event-set}
  \end{subfigure}
\begin{subfigure}[l]{0.23\textwidth}
  \includegraphics[width=1.7in]{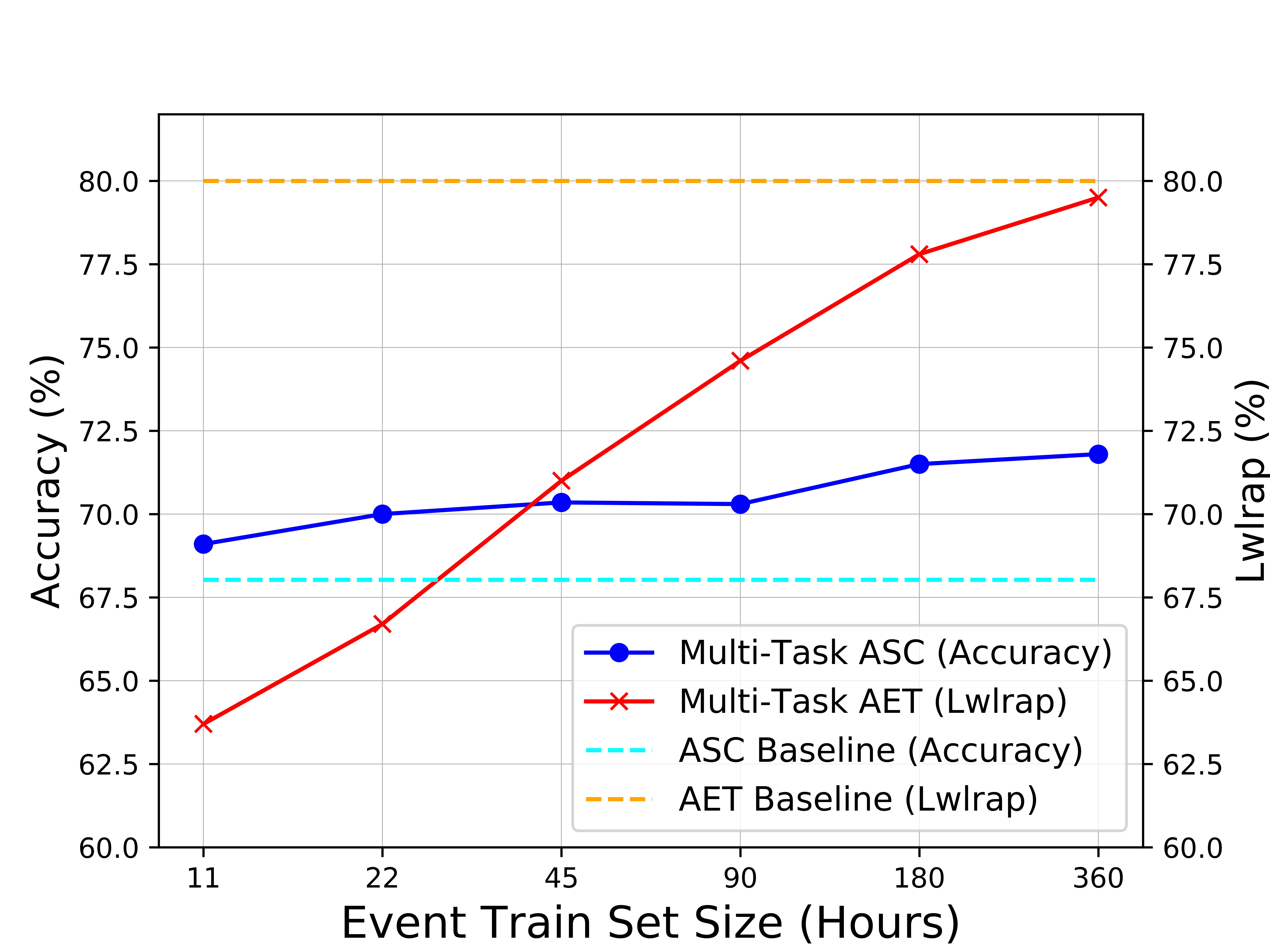}
  \caption{}
  \label{fig: event-set_size}
  \end{subfigure}
  \caption{Studying ASC performance by modifying the AET task:  (a) Events sampled using three different sampling strategies for $k=35, T=170$ hours. (b) AET training data sets of different sizes ($T$) for $k=35$. Events are randomly-sampled. }
  \label{}
\end{figure}

Figure  \ref{fig: event-set} shows that multi-task learning model using the AET event-classes obtained from either of the sampling strategies outperforms the baseline ASC system. It is noteworthy that this improvement in performance is uncorrelated with the choice of the events used to train the AET task. 

\subsection{Impact of AET Dataset Size on ASC Performance}
\label{sec:AET Training Data Size}
To compare the multi-task learning model performance when trained on AET data sets of different sizes ($T$), we randomly sample $k=35$ events and vary the amount of training data corresponding to these events, ranging from 11 to 360 hours. The AET validation data set is kept fixed across all models for fair comparison between the AET performance. Figure  \ref{fig: event-set_size} illustrates that the multi-task learning model outperforms the baseline ASC system for all values $T$. Interestingly, ASC accuracy continues to increase with an increase in the event training data size. The AET data set size and ASC accuracy have a Pearson Correlation Coefficient \cite{oppenhein1999discrete} $ \rho=0.89$, suggesting a strong positive correlation between the ASC performance and the size of the AET data set. As expected, the lwlrap also increases with an increase in training data for AET.
\subsection{Impact of Number of Evens in AET Dataset on ASC Performance}
\label{sec:Choice of Number of Events in the AET Task}
\begin{figure}[h]
  \centering
  \includegraphics[width=0.95\columnwidth]{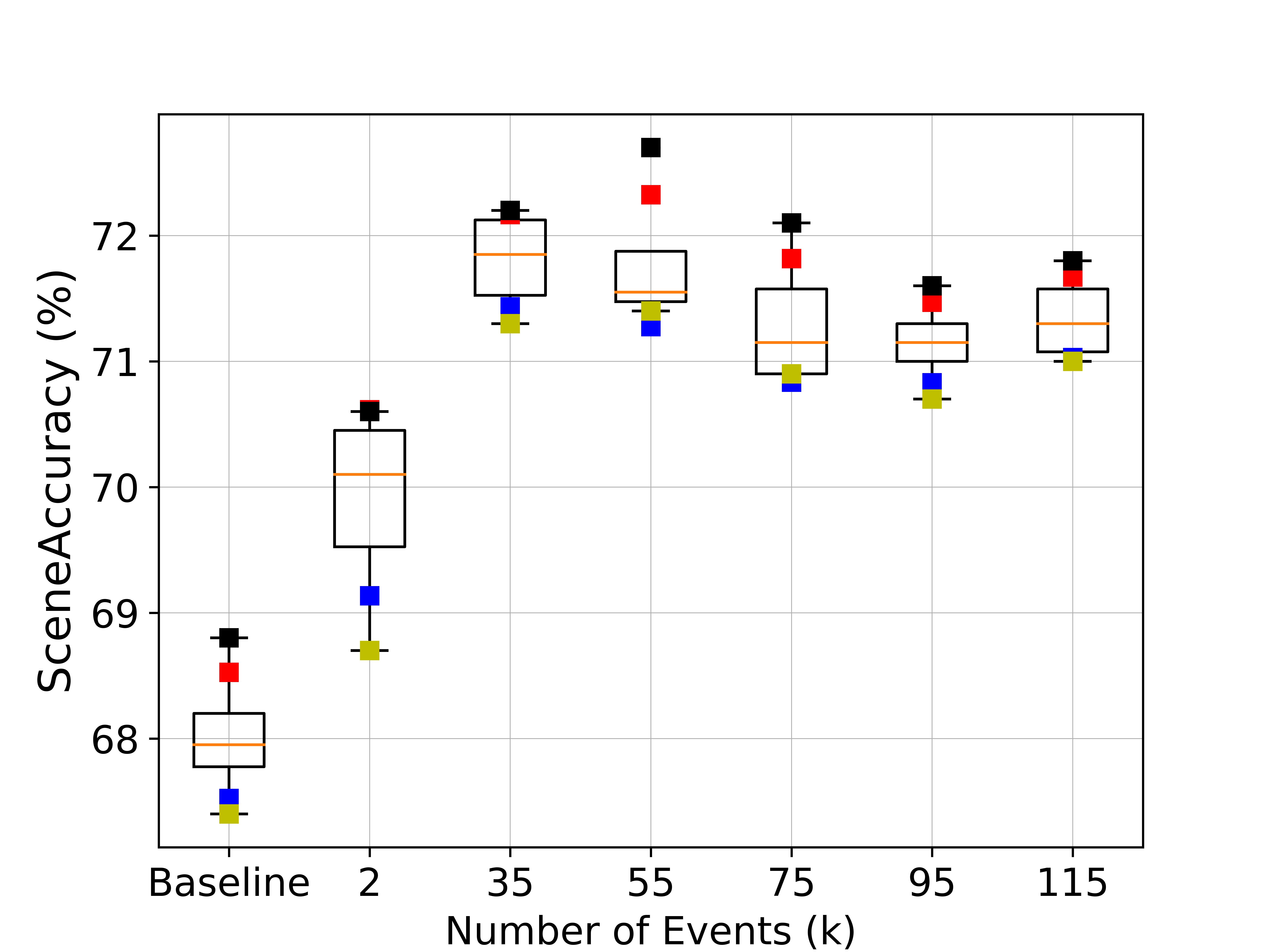}
  \caption{Joint-Training with different number of event-classes ($k$) in the AET data-set. $T$ is 180 hours in all experiments.}
  \label{fig: num_event}
  \vspace*{-2mm}
\end{figure}

To analyze the impact of the number of events ($k$) in the AET data set on ASC performance, we prepare AET data sets with $k$ ranging from 2 to 115, as illustrated in Figure  \ref{fig: num_event} for $T$=180 hours. The event-classes are randomly sampled for $k>2$. For $k=2$, we train the AET task using the event class `speech' and `music', which are most abundant in AudioSet. We observe that even when the AET task is trained to classify only between `music' and `speech' events, the multi-task learning scene classifier outperforms the baseline model. We also observe that after $k=35$, increasing the number of event classes further has marginal influence on the scene accuracy. In general, the ASC performance has a weak positive correlation with the number of events in the AET data sets, $\rho=0.24$.
\vspace*{-2mm}
\subsection{Comparing Pre-Training with Joint-Training} 
\begin{figure}
  \centering
  \includegraphics[width=0.95\columnwidth]{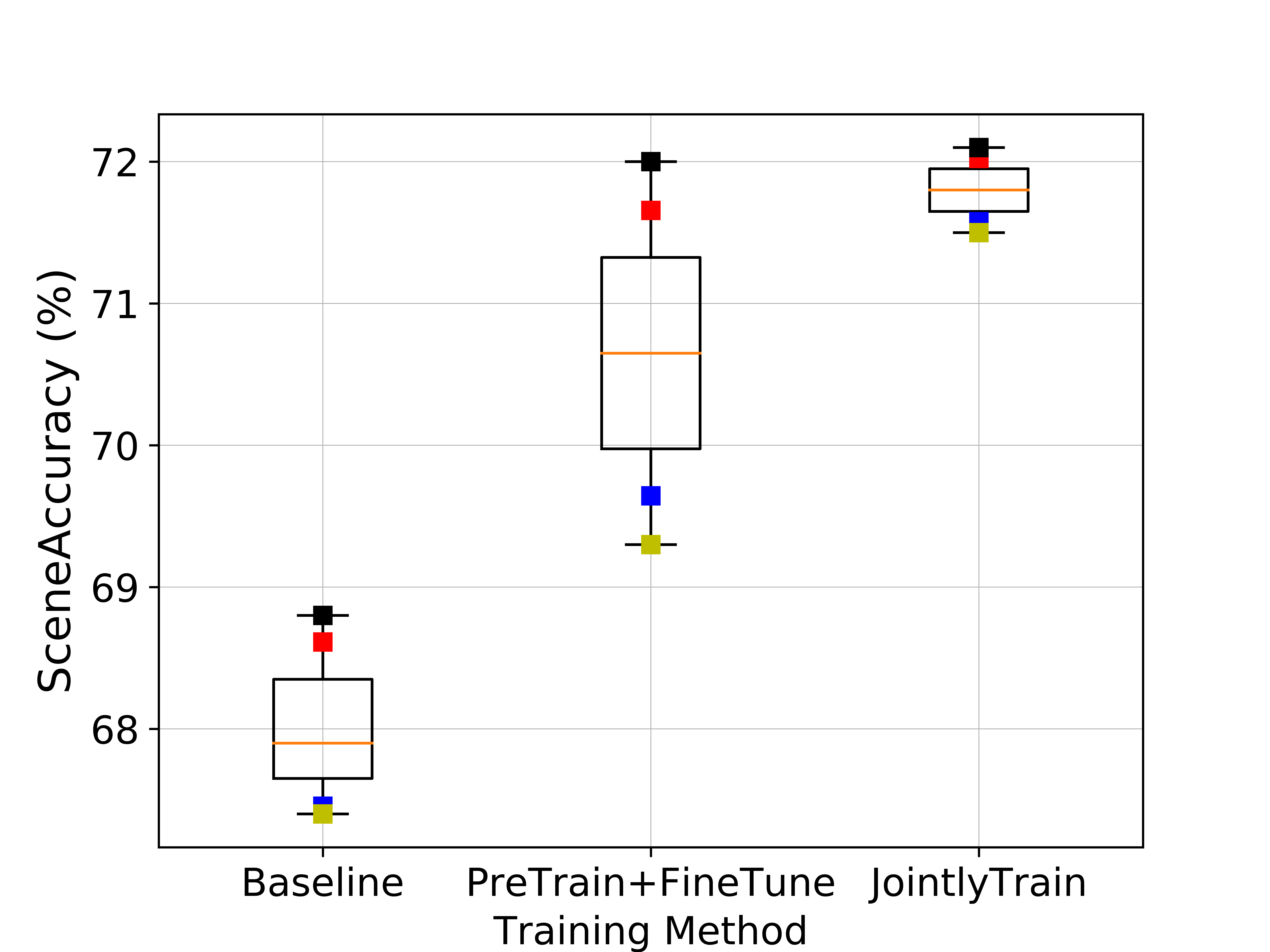}
  \caption{Comparison Between Pre-Training and Multi-Task Learning Frameworks}
  \label{fig: pre-training}
\end{figure}

An alternate framework to leverage AET to improve ASC is to pre-train a network to perform AET and then fine-tune it for ASC \cite{zhang2019cross}. In Figure  \ref{fig: pre-training}, we observe that both frameworks outperform the baseline and joint-training is marginally better than fine tuning. Additionally, joint-training is more scalable and can be simultaneously used for both tasks, while the pre-training is sequential and eventually can be used only for ASC.  
\vspace*{-2mm}
\section{Conclusion}
\label{sec:Conclusion}
We demonstrate empirically that scene classification performance can be consistently improved using event tagging as an auxiliary task in a multi-task learning framework when the individual task losses contribute equally to the overall loss. Through ablation studies on carefully handcrafting the data used for the auxiliary AET task, we show that the performance of the main ASC task improves with the size of the AET data set while being agnostic to the specific choice of events or the number of events used in the AET task. Contrary to previous works, we show empirically that this improvement in ASC performance cannot be attributed to the network gaining knowledge of events or its ability to discern between events. Instead, it is most likely due to the regularization effect of using AET as an auxiliary task. This work opens the door to deeper analysis on the interplay between ASC and AET.
\vspace*{-2mm}

\section{Acknowledgements}
We thank Viktor Rozgic, Chieh-Chi Kao and Qingming Tang at Amazon Inc. for their feedback and suggestions on the experiment design.



\bibliographystyle{IEEEtran}

\bibliography{main}

\end{document}